\documentclass[10pt,a4paper,twoside]{article}
\usepackage{epsfig}
\usepackage{baltlat6}
\usepackage{array}
\usepackage{amssymb}
\begin{document}
\ \
\vspace{0.5mm}
\setcounter{page}{1}
\vspace{8mm}

\titlehead{Baltic Astronomy, vol.\,18, 305--310, 2009}

\titleb{GAMMA-RAY BURST CLASSES FOUND IN THE RHESSI DATA SAMPLE}

\begin{authorl}
\authorb{J. \v{R}\'{\i}pa}{1},
\authorb{C. Wigger}{2},
\authorb{D. Huja}{1} and
\authorb{R. Hudec}{3}
\end{authorl}

\begin{addressl}
\addressb{1}{Charles University,
Faculty of Mathematics and Physics, Astronomical\\
Institute, V Hole\v{s}ovi\v{c}k\'ach 2, 180 00 Prague 8, Czech
Republic;\\ ripa@sirrah.troja.mff.cuni.cz}
\addressb{2}{Kantonsschule Wohlen, Allmendstrasse 26,
5610 Wohlen, Switzerland}
\addressb{3}{Astronomical  Institute, Academy of Sciences of the Czech
Republic,\\
CZ-251 65 Ond\v{r}ejov, Czech Republic}
\end{addressl}

\submitb{Received: 2009 July 20; accepted: 2009 December 1}

\begin{summary}
A sample of 427 gamma-ray bursts (GRBs), measured by the RHESSI
satellite, is studied statistically to determine the number of GRB
groups.  Previous studies based on the BATSE Catalog and recently on the
Swift data claim the existence of an intermediate GRB group, besides the
long and short groups.  Using only the GRBs durations $T_{90}$ and
$\chi^2$ or F-test, we have not found any statistically significant
intermediate group.  However, the maximum likelihood ratio test,
one-dimensional as well as two-dimensional hardness vs.  $T_{90}$ plane,
reveal the reality of an intermediate group.  Hence, the existence of
this group follows not only from the BATSE and Swift datasets, but also
from the RHESSI results.
\end{summary}


\resthead{Gamma-ray burst classes in the RHESSI data sample}
{J. \v{R}\'{\i}pa, C. Wigger, D. Huja, R. Hudec}

\sectionb{1}{INTRODUCTION}

Originally it was found (BATSE data:  Kouveliotou et al. 1993;
Konus-Wind data:  Aptekar et al. 1998) that two GRB classes exist:  a
short one of duration $\lesssim 2$~s and a long one $\gtrsim 2$~s.
They have different celestial distributions (Bal\'azs et al. 1998, 1999;
M\'esz\'aros et al. 2000a,b; Litvin at el. 2001; M\'esz\'aros \&
\v{S}to\v{c}ek 2003; Vavrek et al. 2004, 2008), and they contain
two different central engines (Bal\'azs et al. 2003, 2004; Fox et al.
2005).  Some articles deal or even point to the existence of three GRB
classes in the BATSE database (Horv\'ath 1998, 2002; Mukherjee et al.
1998; Belousova et al. 1999; Balastegui et al. 2001; Horv\'ath et al.
2004, 2006; Chattopadhyay et al. 2007).  Some authors conclude that the
third class (of intermediate duration), observed by BATSE, is a bias
caused by an instrumental effect, placing the intermediate group as a
separate source population in doubt (Hakkila et al. 2000, 2004;
Rajaniemi et al. 2002).  On the other hand, recent papers by Horv\'ath
et al.  (2008), Horv\'ath (2009), and Huja et al.  (2009) indicate that
there is a statistically significant intermediate group in the Swift and
BeppoSAX datasets.

The purpose of this paper is to investigate the number of GRB groups in
a dataset provided by the RHESSI satellite\footnote{
\texttt{~http://hesperia.gsfc.nasa.gov/hessi}} which consists of 427
GRBs covering the period of 2002--2008.  This issue is examined in more
detail in the paper \v{R}\'{\i}pa et al.  (2009).  We analyzed both
one-dimensional distribution of GRB durations and two-dimensional plane
of hardness ratio versus duration.  To determine the number of GRB
groups, standard statistical tests described by Trumpler \& Weaver
(1953), Press et al.  (1992) and Zey et al.  (2006) were used.

\sectionb{2}{DATA SAMPLE}

The Ramaty High Energy Solar Spectroscopic Imager (RHESSI) is a NASA
Small Explorer satellite designed to study hard X-rays and gamma-rays
from solar flares (Lin et al. 2002).  Its spectrometer consists of nine
Ge detectors (Smith et al. 2002) which are only lightly shielded,
ensuring that RHESSI is also useful for detecting non-solar photons from
any direction (Smith et al. 2003).  The energy range for GRB detection
extends from about 30~keV to 17~MeV.  The effective area is around
150~cm$^2$ (Wigger et al. 2006).  We used the RHESSI GRB List\footnote
{\texttt{~http://grb.web.psi.ch}} and the Cosmic Burst List\footnote
{\texttt{~http://www.ssl.berkeley.edu/ipn3/masterli.html}} to find 487
GRBs between 2002 February 14 and 2008 April 25.  There is no automatic
search routine, only if there is a message from any other instrument of
the Interplanetary Network
\footnote{\texttt{~http://www.ssl.berkeley.edu/ipn3/index.html}}, the
RHESSI data are sear\-ched for a GRB signal.

We chose a subset of 427 events with signal/noise ratio higher than 6
and derived count flux curves (with a time resolution higher than 10\,\%
of the burst duration for the vast majority of our entire dataset) and
count fluences from the rear detector segments (except number R2) of the
spectrometer (Smith et al. 2002) in the energy band from 25~keV to
1.5~MeV.

\sectionb{3}{DURATION DISTRIBUTION}

First, we studied one-dimensional duration distribution of $T_{90}$
(Kouveliotou et al. 1993; Fishman et al. 1994).  The histogram consists
of 19 bins on a decimal logarithmic scale (Figures~1 and 2).  The
uncertainty of $T_{90}$ consists of two components.  One is given by the
count fluence uncertainty during $T_{90}$ ($\delta t_{\rm s}$), which is
given by Poissonian noise, and the second one is the time resolution of
derived light curves ($\delta t_{\rm res}$).  The total $T_{90}$
uncertainty $\delta t$ was calculated to be $\delta t = \sqrt{ \delta
t_{\rm s}^2 + \delta t_{\rm res}^2 }.$ We follow the method used by
Horv\'ath (1998) and fitted one, two and three log-normal functions and
used the $\chi^2$ test to evaluate these fits.  The number of GRBs per
bin is at least 5 (except the last bin).  The fit with one log-normal
function is highly unacceptable, because $\chi^2 \simeq 157$ for 17
degrees of freedom ({\it dof}).  The fit with two log-normal functions
({\it dof} = 14) gives $\chi^2=19.13$ which imply a goodness-of-fit
({\it gof}) of 12\%.  The fit with three log-normal functions ({\it dof}
= 11) gives $\chi^2=10.30$ which imply {\it gof} = 41\%.  The assumption
of two groups being represented by two log-normal fits is acceptable,
the fit with three log-normal functions even more.  The question is
whether the improvement in $\chi^2$ is statistically significant.  To
answer this question, we used the F-test, as described by Band et al.
(1997).  The F-test gives a probability of 6.9\% of the improvement in
$\chi^2$ being accidental.  This value is remarkably low, but not enough
to reject the hypothesis that the two log-normal functions are
sufficient to describe the observed duration distribution.  To determine
how the $T_{90}$ uncertainties affect our result, we randomly selected
one half of the bursts and shifted their durations by the full amount of
their uncertainties to lower values and the second half to higher
values.  We then compiled a histogram and recalculated the best-fit
model parameters, $\chi^2$ and F-test.  Only three cases out of ten
performed simulations gave F-test probability lower than 5\% (namely
0.5, 2.0 and 3.5\%).  Therefore, on average, the improvement in $\chi^2$
is insignificant and we cannot proclaim the acceptance of the three
groups of GRBs by the $\chi^2$ method.

Since the number of GRBs is just slightly higher than 5 for many bins,
we also used the maximum likelihood method (see in Horv\'ath 2002) to
fit two and three log-normal functions.  The difference of the
logarithms of the likelihoods $\Delta \ln L = 9.2$ should be
distributed as a half of the $\chi^2$ for 3 degrees of freedom.
From this value we infer that the introduction of a third
group is statistically significant on the 0.036\% level.  To estimate,
how the $T_{90}$ uncertainties affect our result, we again generated ten
different datasets randomly changing the durations by the full amount of
their uncertainties.  All ten simulations give probabilities which are
in favour that introducing of the third group is accidental, much lower
than 5\% (in fact lower than 1\%).  Thus, the hypothesis of introducing
a third group is acceptable when using a maximum likelihood fit.


\begin{figure}[!tH]
\vbox{
\centerline{\psfig{figure=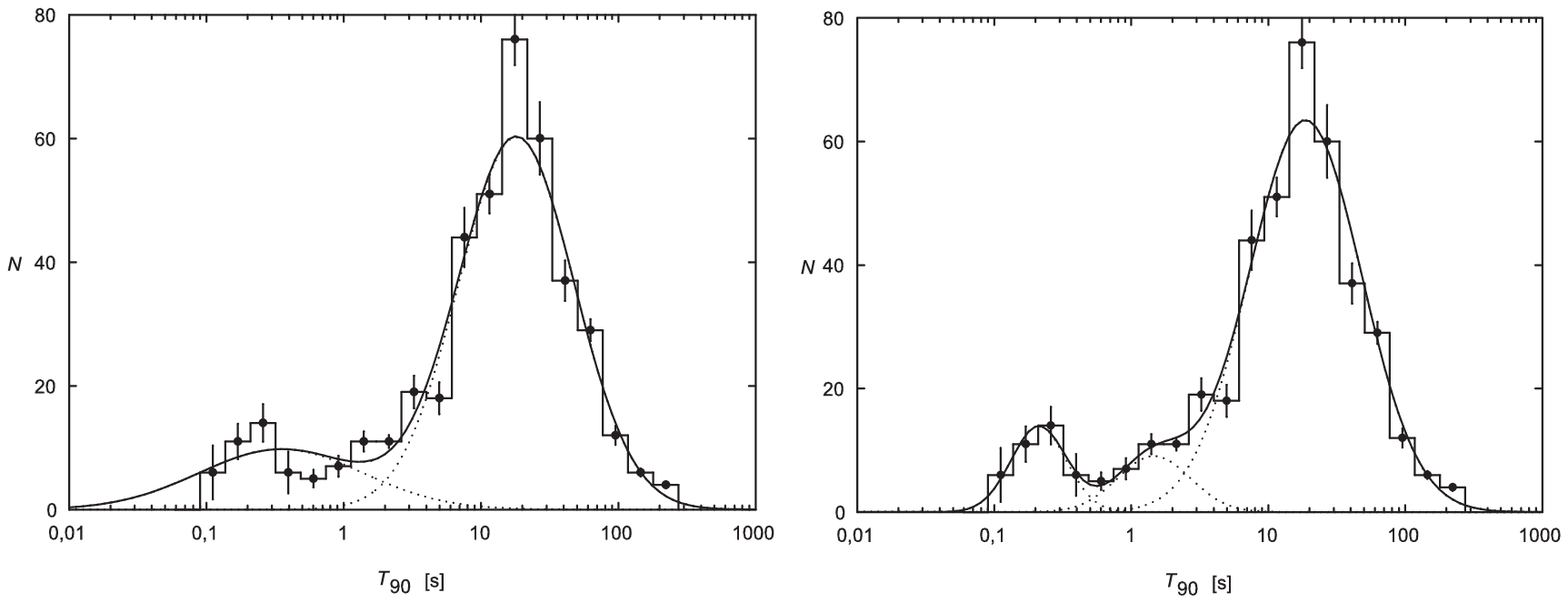,width=124mm,angle=0,clip=}}
\vspace{1mm}
\captionb{1}
{Duration distribution of the 427 RHESSI GRBs (the number of bins is
19).
The best fit with two log-normal functions (left panel) and with three
log-normal functions (right panel).  The error bars are standard
deviations of the number of GRBs per bin for ten different simulated
duration distributions described in the text.}
}
\end{figure}

\sectionb{4}{HARDNESS RATIO VERSUS DURATION}

A two-dimensional scatter plot (duration $T_{90}$ vs. hardness ratio) is
shown in Figure~2.  The hardness ratio is defined as the ratio of two
count fluences $F$ in two different energy bands integrated over the
time interval $T_{90}$.  We used the energy bands (25--120)~keV and
(120--1500)~keV, i.e., $H=F_{120-1500}/F_{25-120}$.  Employing the
maximum likelihood method (see Horv\'ath et al. 2004, 2006 and
references therein), we fit two and three bivariate log-normal
functions.
The difference in the logarithms of the likelihoods $\Delta
\ln L = 10.9$ should be distributed as a half of the $\chi^2$
for
6 degrees of freedom (Horv\'ath et al. 2006) From this value  we infer
that
the introduction of a third group is statistically significant on the
0.13\% level.
To estimate,
how the uncertainties of GRB durations and hardness ratios
effect our result,
we generated ten different datasets randomly changed in durations and
hardness ratios by the full amount of their uncertainties and
recalculated the likelihoods.  All ten simulations infer probabilities,
of introducing the third group being accidental, lower than 5\,\%.
Thus, the hypothesis of introducing the third group is again
acceptable.


\begin{figure}[!tH]
\vbox{
\centerline{\psfig{figure=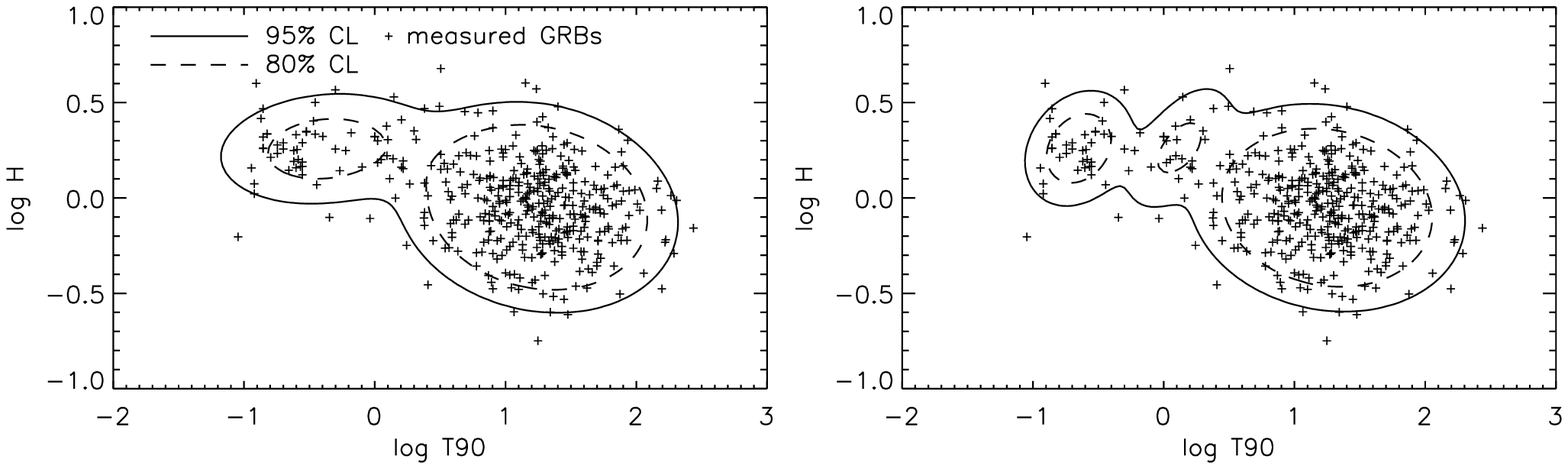,width=124mm,angle=0,clip=}}
\vspace{1mm}
\captionb{2}
{Hardness ratio vs. $T_{90}$ of the RHESSI
GRBs with the best fit of two and three bivariate log-normal functions.}
}
\end{figure}

\sectionb{5}{DISCUSSION AND CONCLUSIONS}

The existence of an intermediate class from the RHESSI $T_{90}$
distribution in not confirmed using the $\chi^2$ method.  However, the
maximum likelihood ratio test on the same data reveals that the
introduction of the third class is statistically significant.  The
hardness ratio versus duration plot for the RHESSI sample also
demonstrates the existence of the third class.  The typical durations
are similar to those obtained with the one-dimensional analysis.  The
typical durations found for BATSE (Horv\'ath et al. 2006) are roughly a
factor of 2 longer than for RHESSI, but consistent for all three
classes.  The shorter durations of the RHESSI GRBs compared to the BATSE
GRBs can be understood in the following way:  for RHESSI, which is
practically unshielded, the background is high (minimum around 1000
counts per second in the (25--1500)~keV band) and varies by up to a
factor of 3. Additionally, the RHESSI sensitivity declines rapidly below
$\approx$\,50~keV.  Weak GRBs (in the sense of counts per second) and
soft GRBs are not so well observed by RHESSI.  Since GRBs tend to be
softer and weaker with time, they rapidly fall bellow the RHESSI
detection limit, resulting in a shorter duration being inferred.

In the RHESSI dataset we confirm the BATSE result that short GRBs are on
average harder than the long GRBs.  The hardness of the intermediate
class found for the RHESSI data is similar to that of the short GRBs.
This is surprising since the intermediate class in the BATSE data was
found to be the softest.  This discrepancy might by explained by the
different definitions of the hardness.  The hardness $H$ for the RHESSI
data is defined as $H=F_{120-1500}/F_{25-120}$, whereas for the BATSE
data $H=F_{100-320}/F_{50-100}$, where the numbers denote energy in keV.
This means that hardnesses measure different behavior of bursts.  The
situation differs even more significantly if we compare hardnesses in
the Swift and RHESSI databases, because the Swift hardnesses are defined
as $H=F_{100-150}/F_{50-100}$ and $H=F_{50-100}/F_{25-50}$ (Horv\'ath et
al. 2008; Sakamoto et al. 2008).

\thanks{This study was supported by the GAUK grant No. 46307, GA\v{C}R
grants No. 205/08/H005 and 205/08/1207, Research Program MSM0021620860
of the Ministry of Education of the Czech Republic and the INTEGRAL PECS
Project 98023.  Discussions with A. M\'esz\'aros are greatly
acknowledged.}

\References

\refb Aptekar R. L., Butterworth P. S., Cline T. L. et al. 1998, AIPC,
428, 10

\refb Band D. L., Ford L. A., Matteson J. L. et al. 1997, ApJ, 485, 747,
Appendix A

\refb Balastegui A., Ruiz-Lapuente P., Canal R. 2001, MNRAS, 328, 283

\refb Bal\'{a}zs L. G., M\'{e}sz\'{a}ros A., Horv\'{a}th I. 1998, A\&A,
339, 1

\refb Bal\'{a}zs L. G., M\'{e}sz\'{a}ros A., Horv\'{a}th I., Vavrek R.
1999, A\&AS, 138, 417

\refb Bal\'{a}zs L. G., Bagoly Z., Horv\'{a}th I. et al. 2003, A\&A,
401, 129

\refb Bal\'{a}zs L. G., Bagoly Z., Horv\'{a}th I. et al. 2004, Baltic
Astronomy, 13, 207

\refb Belousova I. V., Mizaki A., Roganova T. M., Rosental' I. L. 1999,
Astronomy Reports, 43, 734

\refb Chattopadhyay T., Misra R., Chattopadhyay A. K., Naskar M. 2007,
ApJ, 667, 1017

\refb Fishman G. J., Meegan C. A., Wilson R. B. et al. 1994, ApJS, 92,
229

\refb Fox D. B., Frail D. A., Price P. A. et al. 2005, Nature, 437, 845

\refb Hakkila J., Haglin D. J., Pendleton G. N. et al. 2000, ApJ, 538,
165

\refb Hakkila J., Giblin T. W., Roiger R. J. et al. 2004, Baltic
Astronomy, 13, 211

\refb Horv\'{a}th I. 1998, ApJ, 508, 757

\refb Horv\'{a}th I. 2002, A\&A, 392, 791

\refb Horv\'{a}th I., M\'{e}sz\'{a}ros A., Bal\'{a}zs L. G., Bagoly Z.
2004, Baltic Astronomy, 13, 217

\refb Horv\'{a}th I., Bal\'{a}zs L. G., Bagoly Z. et al. 2006, A\&A,
447, 23

\refb Horv\'{a}th I., Bal\'{a}zs L. G., Bagoly Z., Veres P. 2008, A\&A,
489, L1

\refb Horv\'{a}th I. 2009, Ap\&SS, 323, 83

\refb Huja D., M\'{e}sz\'{a}ros A., \v{R}\'{\i}pa J. 2009, A\&A, 504, 67

\refb Kouveliotou C., Meegan C. A., Fishman G. J. et al. 1993, ApJ, 413,
101

\refb Lin R. P., Dennis B. R., Hurford G. J. et al. 2002, Solar Physics,
210, 3

\refb Litvin V. F., Matveev S. A., Mamedov S. V., Orlov V. V. 2001,
Astronomy Letters 27, 416

\refb M\'{e}sz\'{a}ros A., Bagoly Z., Vavrek R. 2000a, A\&A, 354, 1

\refb M\'{e}sz\'{a}ros A., Bagoly Z., Horv\'{a}th I. et al. 2000b, ApJ,
539, 98

\refb M\'{e}sz\'{a}ros A., \v{S}to\v{c}ek J. 2003, A\&A, 403, 443

\refb Mukherjee S., Feigelson E. D., Jogesh B. G. et al. 1998, ApJ, 508,
314

\refb Press W. H., Teukolsky S. A., Vetterling W. T., Flannery B. P.
1992, {\it Numerical Recipes in C}, Cambridge University Press

\refb Rajaniemi H. J., M\"{a}h\"{o}nen P. 2002, ApJ, 566, 202

\refb \v{R}\'{\i}pa J.,	M\'{e}sz\'{a}ros A., Wigger, C. et al. 2009,
A\&A, 498, 399

\refb Sakamoto, T., Barthelmy S. D., Barbier, L. et al. 2008, ApJS, 175,
179

\refb Smith D. M., Lin R. P., Turin P. et al. 2002, Solar Physics, 210,
33

\refb Smith D. M., Lin R. P., Hurley K. C. et al. 2003, Proc. of SPIE
Vol. 4851, 1163

\refb Trumpler R. J., Weaver H. F. 1953, {\it Statistical Astronomy},
Univ. of California Press

\refb Vavrek R., Bal\'{a}zs L. G., M\'{e}sz\'{a}ros A. et al. 2004,
Baltic Astronomy, 13, 231

\refb Vavrek R., Bal\'{a}zs L. G., M\'{e}sz\'{a}ros A. et al. 2008,
MNRAS, 391, 1741

\refb Wigger C., Hajdas W., Zehnder A. et al. 2006, Nuovo Cimento B,
121, 1117

\refb Zey C et al. 2006, NIST/SEMATECH, e-Handbook of Statistical
Methods,\\ \texttt{http://www.itl.nist.gov/div898/handbook/}

\end{document}